%% file: paper.tex
\documentclass[10pt,conference]{IEEEtran}
\IEEEoverridecommandlockouts

\usepackage{graphicx}
\usepackage[linesnumbered,ruled,vlined]{algorithm2e}
\usepackage{stfloats}
\usepackage{textcomp}
\usepackage{xcolor}
\usepackage{xspace}
\usepackage[nointegrals]{wasysym}
\usepackage{pifont}
\usepackage{multirow}
\usepackage{flushend}
\usepackage{caption}
\usepackage{booktabs}
\usepackage{hyperref}
\usepackage{algpseudocode}
\usepackage{pdfpages}
\usepackage{amssymb}
\usepackage{tabularx}
\usepackage{multicol} 
\usepackage[inkscapelatex=false]{svg}
\hypersetup{
colorlinks = true, 
linkcolor=black, 
citecolor=black, 
urlcolor=black 
}
\usepackage{url}
\usepackage{enumitem}
\usepackage{braket}
\usepackage{amsmath}
\usepackage{caption}
\usepackage[breakable]{tcolorbox}
\usepackage{colortbl}
\usepackage{pifont}
\usepackage{float}                  
\usepackage{subfig}                 
\usepackage{overpic}                
\usepackage{tikz}
\usetikzlibrary{quantikz}
\usetikzlibrary{positioning}
\usetikzlibrary{arrows.meta}
\usepackage{color}
\usepackage{listings}
\usepackage[export]{adjustbox}

\definecolor{bleudefrance}{rgb}{0.19, 0.55, 0.91}

\newcommand\black[1]{\textcolor{black}{#1}}

\input{macros}

\begin{document}
%
\title{M2QCode: A Model-Driven Framework for Generating Multi-Platform Quantum Programs}

\author{\IEEEauthorblockN{Xiaoyu Guo}
\IEEEauthorblockA{
\textit{Kyushu University}\\
guo.xiaoyu.961@s.kyushu-u.ac.jp}
\and
\IEEEauthorblockN{Shinobu Saito*}
\IEEEauthorblockA{
\textit{NTT, Inc. Computer \& Data Science Laboratories}\\
shinobu.saito@ntt.com}
\and
\IEEEauthorblockN{Jianjun Zhao*}
\IEEEauthorblockA{
\textit{Kyushu University}\\
zhao@ait.kyushu-u.ac.jp}
\thanks{* Corresponding author.}
}

\maketitle
\begin{abstract}

With the growing interest in quantum computing, the emergence of quantum supremacy has marked a pivotal milestone in the field. As a result, numerous quantum programming languages (QPLs) have been introduced to support the development of quantum algorithms. However, the application of Model-Driven Development (MDD) in quantum system engineering remains largely underexplored. This paper presents an MDD-based approach to support the structured design and implementation of quantum systems. Our framework enables the automatic generation of quantum code for multiple QPLs, thereby enhancing development efficiency and consistency across heterogeneous quantum platforms. The effectiveness and practicality of our approach have been demonstrated through multiple case studies.
\end{abstract}

\begin{IEEEkeywords}
Model-Driven Development, Automatic Code Generation, Quantum Software
\end{IEEEkeywords}

\section{Introduction}
\label{sec:introduction}

Quantum computing promises exponential computational power to solve complex problems, promoting the development of various quantum devices and simulators~\cite{qiskit2024, svore2018q, cirq2025google, braket}. However, implementing quantum programs across different quantum devices remains a challenge. As quantum hardware and quantum programming languages (QPLs) continue to diversify, adapting programs to specific QPLs has become increasingly labor-intensive. This growing heterogeneity requires approaches that streamline development and reduce manual errors. In particular, a unified code generation strategy is needed to support development across multiple QPLs.

Designing such a cross-QPL code generation framework poses several challenges, including differences in gate representations (e.g., Qiskit~\cite{qiskit2024} uses \textit{CX}, while Cirq~\cite{cirq2025google} uses \textit{CNOT}), syntax, standard libraries, and abstraction levels. Supporting hybrid quantum-classical constructs, such as loops, conditionals, and runtime feedback, also requires careful integration of classical control logic with quantum operations. To address these challenges, we adopt a model-driven approach that enables the automated generation of hybrid quantum-classical systems across multiple QPLs. Our approach transforms high-level system specifications into code for various QPLs, facilitating interoperability across heterogeneous platforms and improving development scalability.

In industrial quantum computing, the demand for cross-QPL code generation is increasingly urgent. In response to gate-level noise and inherent errors in today’s Noisy Intermediate-Scale Quantum (NISQ) quantum hardware, NTT has introduced the concept of N-version programming (NVP), a fault-tolerant strategy adapted from classical software engineering~\cite{10771417}. This technique involves generating and executing multiple semantically equivalent versions of a quantum program, called N-version quantum software systems (NVQS), across heterogeneous hardware and QPL platforms. By aggregating the resulting probability distributions, it becomes possible to derive high-confidence measurement outcomes. However, generating NVQS implementations is challenging, as manual translation between QPLs is error-prone and time-consuming, while existing tools lack support for multi-language output, making them inadequate for industrial deployment.

For example, the Quantum UML Profile~\cite{perez2022design} extends UML to model quantum elements, such as qubits, gates, and measurements, allowing visual modeling and translation into Qiskit code. Another example is the Qiskit Code Assistant~\cite{dupuis2024qiskit}, which uses large language models (LLMs) fine-tuned for quantum computing to assist Qiskit SDK~\cite{fingerhuth2018open} users. Although promising, these tools are limited to Qiskit and do not support multi-QPL or cross-platform code generation. Qiskit Code Assistant cannot automatically generate complete and correct quantum programs; it typically serves as an auxiliary aid rather than a fully automated solution. As such, they fall short of addressing the broader need for automated, multi-QPL code generation required for scalable NVP-based quantum software systems.

To overcome these limitations, we propose M2QCode, a framework for generating multi-platform quantum programs. \black{M2QCode supports two key features: \textit{Multiple Quantum Code Generation}, which generates semantically equivalent programs in different QPLs such as Qiskit~\cite{qiskit2024}, Cirq~\cite{cirq2025google}, Q\#~\cite{svore2018q}, and Braket~\cite{braket}; and \textit{Hybrid Quantum System Generation}, which produces hybrid quantum-classical systems.} By combining MQCG and HQSG, M2QCode offers greater flexibility and broader platform support than existing Qiskit-specific tools. \black{We evaluated M2QCode on a quantum system industry dataset collected by Qiskit~\cite{jimenez2024code}}, demonstrating its ability to produce functionally equivalent and executable programs on multiple platforms.

The main contributions of this paper are as follows.

\begin{itemize}
    \item We present M2QCode, a model-driven code generation framework to automatically generate quantum programs in multiple quantum programming languages, including Qiskit, Cirq, Q\#, and Braket.
    
    \item We introduce support for hybrid quantum-classical systems through Hybrid Quantum System Generation (HQSG), enabling the generation of both classical and quantum components.
    
    \item \black{We evaluated M2QCode through seven industry case studies collected by Qiskit~\cite{jimenez2024code}, demonstrating its ability to model and generate functionally equivalent and executable quantum systems on diverse target platforms.}
\end{itemize}

The remainder of this paper is structured as follows. Section~\ref{sec:background} provides background on quantum computing and code generation. Section~\ref{sec:methodology} describes our approach. Section~\ref{sec:evaluation} presents the experimental setup and results. Section~\ref{sec:discussion} discusses our findings. The related work is reviewed in Section~\ref{sec:related work}, and the paper is concluded in Section~\ref{sec:conclusion}.

\section{Background}
\label{sec:background}

This section introduces the basic concepts of quantum computing, followed by a brief overview of code generation techniques relevant to quantum software development.

\subsection{Quantum Computing}

\subsubsection{Quantum Bit (Qubit)}
A quantum bit, or qubit, is the fundamental unit of information in quantum computing. Unlike a classical bit, a qubit can exist in a superposition of states $|0\rangle$ and $|1\rangle$, with specific probabilities of collapsing to either state upon measurement. Mathematically, a qubit is expressed as:

\begin{equation}
|\psi\rangle = \alpha|0\rangle + \beta|1\rangle,\ \text{where}\ |\alpha|^2 + |\beta|^2 = 1
\label{equation:quantum_state}
\end{equation}

In equation~\ref{equation:quantum_state}, $\alpha$ and $\beta$ are the probability amplitudes for states $|0\rangle$ and $|1\rangle$, respectively. The likelihood of measuring the qubit in each state is given by $|\alpha|^2$ and $|\beta|^2$, with the total probability summing to 1.

\subsubsection{Quantum Gates and Circuits}
Quantum operations are represented by logic gates, unitary operators that transform qubit states according to their corresponding unitary matrices. For $N$ qubits, a quantum gate is described by a matrix of $2^N \times 2^N$. These gates serve various roles: the Pauli gates perform rotations, the Hadamard gate introduces superposition, and the CNOT gate creates controlled operations between qubits. Some gates operate on individual qubits, while others act on multiple qubits.

Recent advances in quantum hardware have enabled the development of dynamic circuits~\cite{baumer2024efficient}, which support classical processing during the execution of a quantum circuit. This allows for mid-circuit measurements and feedforward operations, where measurement outcomes influence subsequent gate operations. While such capabilities enhance algorithm efficiency, they also introduce additional complexity to the quantum code generation.

\subsubsection{Quantum Programming Framework}
\label{subsec:qiskit}

Several open-source frameworks have been developed to support quantum programming. Among the most widely used are Qiskit~\cite{qiskit2024}, Q\#~\cite{svore2018q}, Cirq~\cite{cirq2025google}, and Braket~\cite{braket}, which serve as the primary target languages supported by our framework.

\begin{itemize}

\item Qiskit is a Python-based SDK that supports the creation, simulation, and execution of quantum circuits. It also offers a dynamic circuit suite for integrating classical control logic into quantum programs.

\item Cirq is another Python-based library that provides fine-grained control over circuit construction, enabling optimization for specific hardware constraints.

\item Q\# is a domain-specific language developed by Microsoft as part of the Quantum Development Kit (QDK). It features strong typing and a rich development environment tailored for scalable quantum software.

\item Braket, developed by Amazon Web Services, is a managed quantum service with a Python-based SDK for building, simulating, and executing quantum circuits.
\end{itemize}

The diversity of QPLs presents challenges for code generation, particularly in designing transformation rules that accommodate variations in syntax, semantics, and runtime capabilities.

\subsection{Code Generation}

Code generation~\cite{kelly2008domain} refers to the automated creation of source code from higher-level representations, such as models~\cite{usman2008ujector}, specifications~\cite{moreira2010automatic}, or intermediate languages~\cite{draves2005compiler}. It aims to improve developer productivity, reduce errors, and ensure consistency across software systems.
Code generation methods are generally classified into traditional approaches~\cite{usman2008ujector,moreira2010automatic,draves2005compiler} and machine learning (ML) based approaches~\cite{li2022competition,liu2023your}. Traditional methods generate code from structured artifacts, such as domain-specific languages (DSLs), UML models, or formal specifications. ML-based methods, including large language models (LLMs), produce code from natural language descriptions with minimal human input~\cite{liu2023your}.

Although code generation is well established in classical software engineering, applying these techniques to quantum computing poses new challenges. Traditional approaches require the design of quantum-specific abstractions and transformation rules for various QPLs. ML-based methods can generate simple quantum programs, but often lack the domain knowledge needed to handle complex and customized quantum algorithms~\cite{vishwakarma2024qiskit}.

\section{Quantum System Modeling}
\label{sec:modeling}

In this paper, we adopt QuanUML~\cite{guo2025quanuml} as our modeling language. This section introduces the key modeling constructs defined in the QuanUML specification that are used throughout the remainder of the paper.

\subsection{Quantum System Modeling}
The class diagram in high-level modeling provides a structural view of hybrid quantum-classical systems. The system is represented as a set of modules, each containing classical components and quantum algorithm components. In the extended class diagram, the entire system is modeled as a parent class, with its components represented as child classes based on the system design. The functions are grouped within the appropriate child classes. For quantum components, a complete quantum algorithm is modeled as a child class and further detailed in the sequence diagram. To distinguish between classical and quantum components, the stereotype $\langle\langle \textbf{Quantum} \rangle\rangle$ is applied to quantum-related classes, while classical components do not require additional labels.

\subsection{Quantum Circuit Modeling}
\begin{figure}
    \centering
    \includegraphics[width=0.9\linewidth]{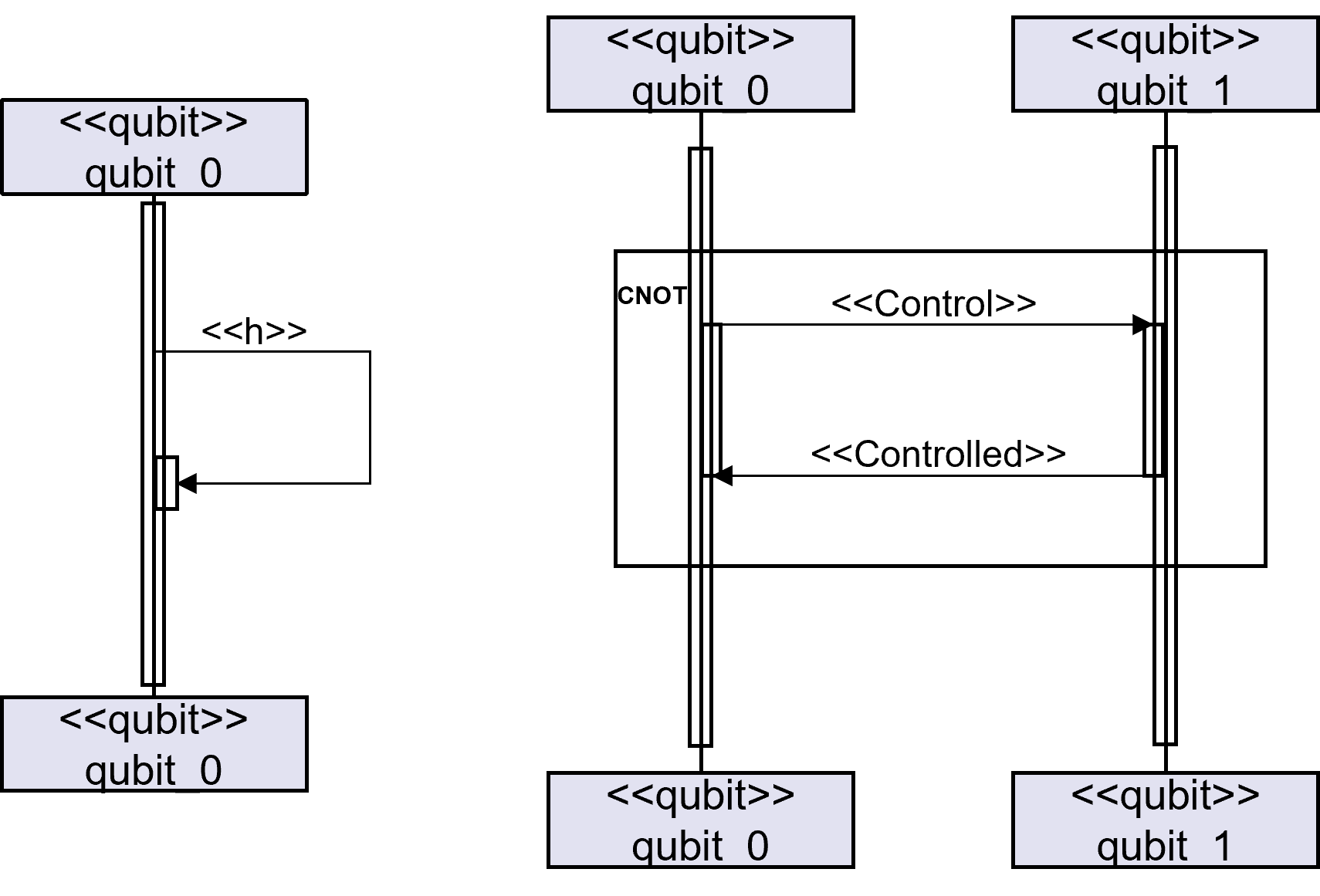}
    \caption{Modeling specifications for quantum gates in UML (adapted from~\cite{guo2025quanuml}).}
    \label{fig:gate}
\end{figure}

\subsubsection{Qubits} In the quantum UML specification, qubits are modeled as distinct entities and annotated with the stereotype $\langle\langle \textbf{qubit} \rangle\rangle$, distinguishing them from classical variables labeled $\langle\langle \textbf{classical\_bit} \rangle\rangle$. The qubit lifeline represents its state from initialization to collapse.

\subsubsection{Gate Operations} Quantum gates introduce superposition and entanglement, making them critical yet challenging to model. Figure~\ref{fig:gate} shows the modeling of the Hadamard and CNOT gates. \black{The left side shows single-qubit gates (e.g., Hadamard) are} modeled as asynchronous messages because of the absence of control flow, and their lifelines must complete before the next operation. \black{The right side shows multi-qubit gates (e.g., CNOT) are modeled as synchronous grouped messages to reflect control relationships and phase kickback.} The control flow is labeled with $\langle\langle \textbf{control} \rangle\rangle$, while the target qubit may be labeled with $\langle\langle \textbf{controlled} \rangle\rangle$, or omitted when unnecessary. This approach enables intuitive visualization of internal gate relationships. Some gates, such as the Swap gate, require special treatment.

\subsubsection{Quantum Measurement} Measurement collapses the state of a qubit into a classical bit, serving as an interface between quantum and classical components. In the UML model, this is represented by an asynchronous message from the qubit to the classical bit, with the lifeline of the qubit terminating at measurement.

\begin{figure}
    \centering
    \includegraphics[width=0.9\linewidth]{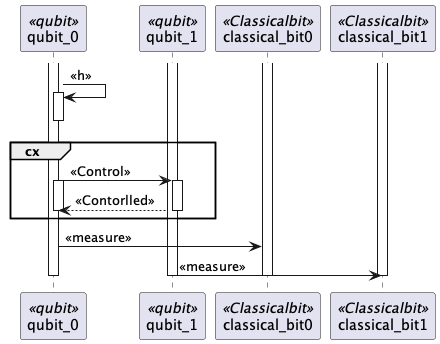}
    \caption{UML modeling specification of the Bell State circuit (adapted from~\cite{guo2025quanuml}).}
    \label{fig:bell_state}
\end{figure}

\textit{Example: Bell State Circuit.} Figure~\ref{fig:bell_state} shows a UML specification of a Bell State circuit using PlantUML~\cite{plantuml}. The circuit uses two qubits to generate entanglement, and two classical bits to store measurement results. The Hadamard gate introduces superposition to \texttt{qubit\_0}, and the \texttt{cx} (CNOT) gate entangles \texttt{qubit\_0} and \texttt{qubit\_1}. The control relationship indicates that \texttt{qubit\_1} is controlled by \texttt{qubit\_0}. Measurement collapses the qubit states, terminates their lifelines, and stores the outcomes in classical bits.

\section{Approach}
\label{sec:methodology}

\begin{figure*}
    \centering
    \includegraphics[width=0.95\linewidth]{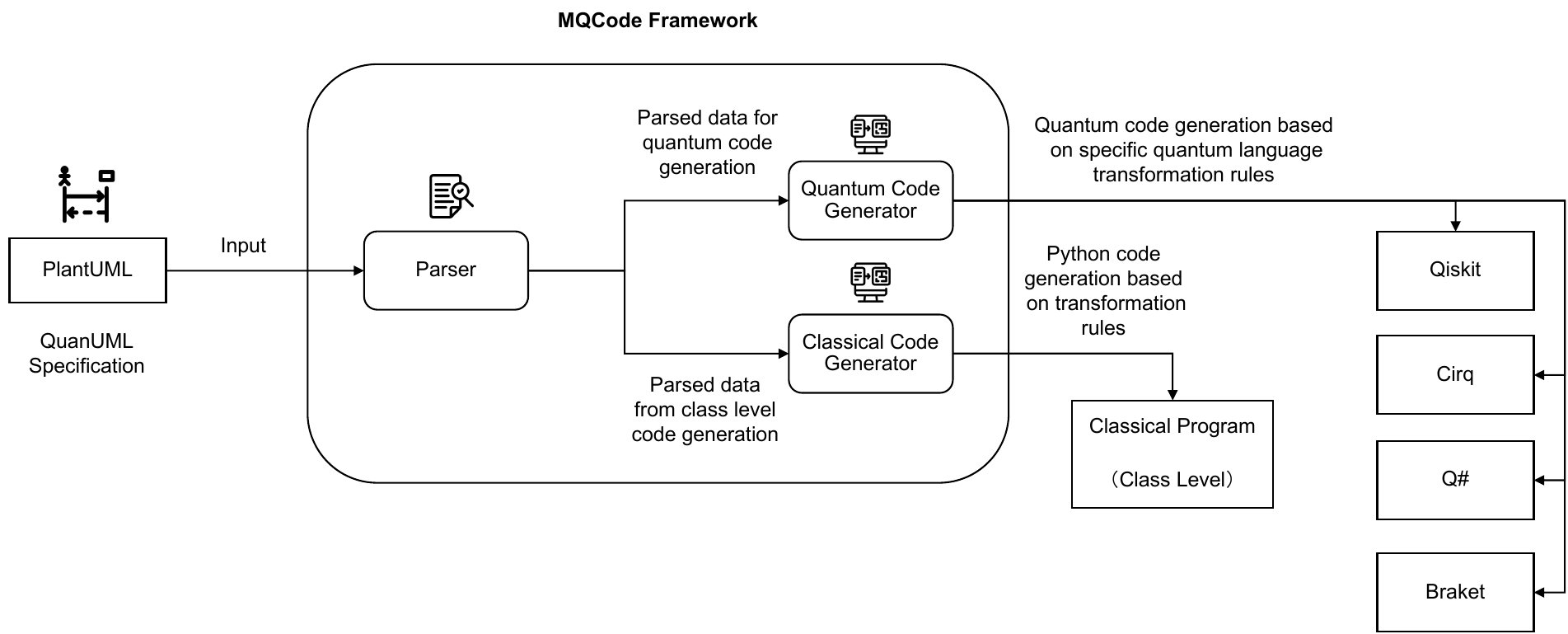}
    \caption{M2QCode Workflow}
    \label{fig:workflow}
\end{figure*}

Model-Driven Development (MDD) relies on model transformations to support efficient design, analysis, and implementation of systems, reducing the manual coding effort. In this paper, we focus on transforming a Platform-Independent Model (PIM) into executable quantum programs across multiple quantum programming languages. The general transformation process is illustrated in Figure~\ref{fig:workflow}, which consists of two main steps: (1) extracting information from the PIM into an intermediate representation (Section~\ref{subsec:PIM2IR}), and (2) generating quantum and classical code from the intermediate representation (Section~\ref{subsec:IR2Code}).

The PIM is constructed using an extended UML-based modeling approach and serves as the abstract input model. Custom transformation algorithms are defined to map both structural and behavioral elements from the PIM to an intermediate representation (IR), and subsequently to platform-specific quantum and classical code.

\subsection{Transformation from a Platform-Independent Model to an Intermediate Representation}
\label{subsec:PIM2IR}

The quantum PIM serves as an input to the transformation process, and the intermediate representation stores the extracted structural and behavioral information. The PIM includes an extended UML class diagram for capturing the system structure and a sequence diagram for describing the behavior of the quantum algorithm.

The class diagram represents the system at a high level, whereas the sequence diagram captures gate-level quantum operations. We developed a parser and internal data representation to extract and store the information from both diagrams. Specifically, the extracted class diagram data include packages, classes, operations, and attributes. The sequence diagram data includes qubit declarations, gate operations, and measurements, all encapsulated as UMLBlocks in the intermediate representation.

\begin{algorithm}[ht]
\small
\caption{Transform Quantum PIM to Intermediate Representation}
\label{alg:parser}
\KwIn{Quantum PIM Class Diagram and Optional Sequence Diagram}
\KwOut{Intermediate Representation}

\While{package is not empty}{
    Element $\leftarrow$ \texttt{GetNextElement(ClassDiagram)}\;

    \uIf{Element is Package}{
        \texttt{TransformPackage}(Element)\;
    }
    \uElseIf{Element is Class}{
        \texttt{TransformClass}(Element)\;
        \texttt{TransformOperationsAndProperties}(Element)\;
    }
}

\texttt{TransformAssociations}(ClassDiagram)\;

\uIf{SequenceDiagram is specified}{
    ElementList $\leftarrow$ \texttt{GetAllElements(SequenceDiagram)}\;

    \texttt{TransformQubitsAndClassicalBits}(ElementList)\;
    \texttt{TransformQuantumGates}(ElementList)\;
    \texttt{TransformMeasurements}(ElementList)\;
}

\Return IntermediateRepresentation\;
\end{algorithm}

The algorithm~\ref{alg:parser} outlines the PIM-to-IR transformation process.

\subsubsection{Class Diagram} Each package in the UML class diagram is recursively transformed into a structured object in the IR. Classes are extracted along with their operations and attributes. Associations among classifiers are also converted into corresponding association objects. Quantum-related classes are annotated with the stereotype $\langle\langle \textbf{Quantum} \rangle\rangle$ during transformation.

\subsubsection{Sequence Diagram} If a sequence diagram is provided, each Actor is converted into a Qubit or ClassicalBit object based on its stereotype. Combined fragments (e.g., groups) and messages are transformed into multi-qubit operation objects, while conditional fragments (e.g., Alt) become conditional operation blocks. Asynchronous messages correspond to single-qubit operations, with operation types inferred from stereotypes.

\subsection{Transformation from an Intermediate Representation to Quantum Code}
\label{subsec:IR2Code}

\subsubsection{Quantum Code Generation}
The first step in code generation produces a quantum program based on IR for a given target QPL. The algorithm~\ref{alg:quantumCodeGen} summarizes the procedure.

\begin{algorithm}[ht]
\small
\caption{Quantum Code Generation from Intermediate Representation}
\label{alg:quantumCodeGen}

\KwIn{Intermediate Representation (IR), Target Quantum Language}
\KwOut{Executable Quantum Program}

TargetLanguage $\leftarrow$ \texttt{GetTargetQuantumLanguage()}\;
\texttt{ImportQuantumPackage}(TargetLanguage)\;

\texttt{CreateQubitsAndClassicalBits}(IR)\;
\texttt{CreateGateOperations}(IR)\;
\texttt{CreateExecutionInstructions}(TargetLanguage)\;

\Return QuantumProgram\;
\end{algorithm}

The process starts by importing the necessary libraries, followed by creating qubit and classical bit variables. Some QPLs automatically allocate these resources, while others require explicit declarations.

The \textit{QuantumCircuit} object is then defined, and the quantum gate and measurement operations are added in sequence. Single-qubit gates contain one target qubit as a parameter, while multi-qubit gates and measurement operations contain multiple parameters. Conditional operations (e.g., dynamic circuits) are represented as operations along with the classical condition and are only supported in select QPLs. Finally, simulation settings and execution instructions (e.g., number of shots) are appended to complete the executable program.

\subsubsection{Classical Code Generation}
The second step involves generating Python code for the classical components of the IR, \black{due to major QPL frameworks relying on Python}. Although UML-to-code tools exist, they typically require modification to accommodate the characteristics of quantum systems.

The process begins by importing the previously generated quantum modules. Then elements such as \texttt{Class}, \texttt{Interface}, and \texttt{Abstract} are converted into Python class definitions, following the directory structure defined in the IR. Attributes and functions are mapped accordingly, though function bodies remain empty by design, allowing developers to implement logic manually.

\subsection{Supporting Tools}

The M2QCode framework comprises both modeling and transformation components. The modeling process uses PlantUML~\cite{plantuml}, a lightweight tool that generates UML diagrams through simple text-based descriptions, enabling the quick modeling of quantum systems.
The transformation process uses a custom parser developed with Peggy~\cite{peggy}, a JavaScript-based parser generator. It converts UML models into an intermediate representation.

The quantum code generator is implemented in JavaScript. We define QPL-specific conversion rules using JavaScript, enabling lightweight and flexible generation of target code.

In summary, M2QCode provides an end-to-end workflow for automatically generating quantum circuit code from high-level models. This approach improves code portability, reduces manual development effort, enhances software quality, and extends model-driven development to the quantum computing domain.


\section{Evaluation of Implemented Transformations for Quantum Systems}
\label{sec:evaluation}

\subsection{Research Questions (RQs)}
In our evaluation, we assess M2QCode's ability to model and generate quantum systems across multiple QPLs, guided by the following research questions (RQs):
\begin{itemize}
    \item \textbf{RQ1}: \textit{How does M2QCode model practical quantum systems?} As a foundational aspect of our research, this question asks whether M2QCode can effectively model real-world quantum systems. Addressing it demonstrates the framework's feasibility and applicability in real-world settings, highlighting its potential for adoption in model-driven quantum software development workflows.
    \item \textbf{RQ2}: \textit{How effective is the generation of the quantum system?} This question examines whether the quantum system generated by M2QCode preserves the semantic and structural information of the original quantum UML model. 
    \item \textbf{RQ3}: \textit{Can M2QCode generate functionally equivalent quantum code across multiple QPLs?} This question evaluates whether the quantum code generated by M2QCode is functionally equivalent to the Platform Independent Model (PIM) from which it is derived. Achieving such functional equivalence is a key indicator of the correctness and reliability of the code generation process.
\end{itemize}

\subsection{Case Studies}
To evaluate the performance of M2QCode in modeling and generating quantum systems, we used an open-source dataset of classical quantum systems~\cite{jimenez2024code}, which includes seven curated case studies. Each case satisfies the following criteria: (1) it represents a concrete application of a quantum algorithm, (2) it provides sufficient information to model a complete hybrid classical-quantum system, (3) it follows a gate-based quantum programming paradigm, and (4) it addresses a real-world problem reported by a company or organization.
Table~\ref{tab:cases} presents an overview of the seven selected systems, including their identifiers, the organizations that propose them, the targeted problems, and the quantum algorithms used in each case.

Using our extended UML-based quantum modeling approach, we reformulated these seven quantum system cases to assess M2QCode's capabilities in systematically analyzing the system, extracting semantic information, and automating the generation of executable quantum programs.

\begin{table*}[h!]
\centering
\caption{Overview of Selected Quantum Systems}
\label{tab:cases}
\begin{tabular}{@{}clp{5.6cm}p{5.2cm}@{}}
\toprule
\textbf{ID} & \textbf{Proposer(s)} & \textbf{Domain / Problem} & \textbf{Quantum Algorithm / Method} \\
\midrule
C1 & Airbus & Multidisciplinary design optimization & HHL (Harrow–Hassidim–Lloyd) \\
C2 & Airbus & Computational fluid dynamics (CFD) for aircraft design & Quantum machine learning (QML) \\
C3 & Boehringer & Imaging of molecular structures in biological tissues & Quantum approximate optimization algorithm (QAOA) \\
C4 & BASF, Boehringer & Simulation of chemical reactions and molecular properties & Variational quantum eigensolver (VQE) \\
C5 & Munich Re, SAP & Pallet and truck load optimization in logistics & QAOA \\
C6 & IBM (Qiskit Finance) & Financial modeling and risk analysis & QML \\
C7 & VW, SAP & Routing in autonomous driving and vehicle production & QAOA variant \\
\bottomrule
\end{tabular}
\end{table*}

\subsection{Evaluation Metrics}
To assess the effectiveness of M2QCode in modeling and generating quantum systems, we adopt the following evaluation metrics, aligned with research questions (RQs):

\begin{itemize}
    \item \textit{Precision \& Recall}: To evaluate the effectiveness of M2QCode in generating the classical components of quantum systems (RQ2), we use precision and recall. Precision measures the proportion of correctly generated code segments among all segments produced, while recall measures the proportion of relevant target segments that were successfully generated.

    \begin{equation}
        \text{Precision} = \frac{\text{Relevant elements}}{\text{Relevant elements} + \text{False Positives}}
    \end{equation}
    
    \begin{equation}
        \text{Recall} = \frac{\text{Relevant elements}}{\text{Relevant elements} + \text{False Negatives}}
    \end{equation}

    \item \textit{CodeBLEU}: To evaluate the effectiveness of M2QCode in generating quantum code (RQ2), we adopt the CodeBLEU metric, which evaluates both syntactic and semantic similarity between the generated code and a reference implementation. CodeBLEU extends the traditional BLEU metric by combining the following four components:

    \begin{itemize}
        \item \textbf{Standard BLEU Score}: Measures n-gram overlap between the generated and reference code.
        \item \textbf{Weighted n-gram Match}: Applies weights to code tokens based on their semantic importance (e.g., control keywords, identifiers), prioritizing critical syntax elements.
        \item \textbf{Syntax Match}: Compares abstract syntax trees (ASTs) to capture structural similarity between programs.
        \item \textbf{Semantic Match}: Applies data flow analysis to evaluate whether the generated code preserves the semantics of the reference program.
    \end{itemize}

    This combination makes CodeBLEU particularly well-suited for evaluating the syntactic and semantic correctness of source code in generation tasks.

    \item \textit{Kullback–Leibler (KL) divergence}: To evaluate functional equivalence across QPLs (RQ3), we use KL divergence to measure the difference between the measurement probability distribution \( p \) of the generated program and the reference distribution \( q \). A lower KL divergence indicates a higher functional equivalence, which means that the generated quantum program behaves more consistently with the expected reference output.

    \begin{equation}
        D_{\text{KL}}(p \parallel q) = \sum_{i} p(i) \log \frac{p(i)}{q(i)}
    \end{equation}
\end{itemize}

\subsection{Experiment Setup}
Our experimental setup consists of the following key steps:

\begin{itemize}
    \item We model each hybrid quantum-classical system in the selected case studies using extended UML class and sequence diagrams.
    
    \item We apply M2QCode to automatically generate the corresponding classical Python code and quantum code for multiple QPLs.
    
    \item We evaluate the effectiveness of the overall modeling and code generation process using the predefined static evaluation metrics described above.
    
    \item We verify the generated quantum code across different QPLs by executing it on quantum simulators, depending on their feasibility and availability.
\end{itemize}

\subsection{Evaluation Results and Analysis}

\subsubsection{RQ1: How does M2QCode model practical quantum systems?}

\begin{figure*}
    \centering
    \includegraphics[width=\linewidth]{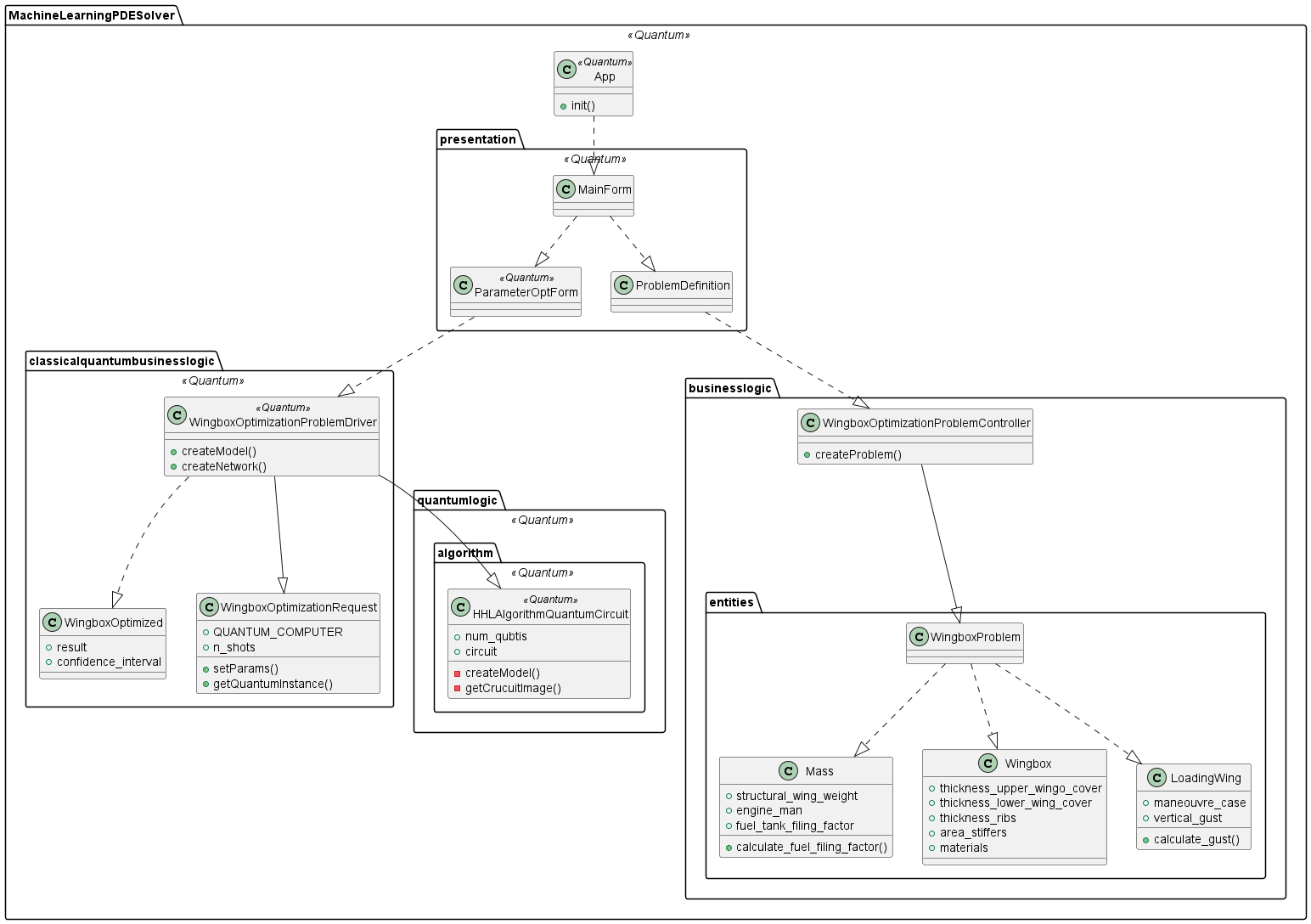}
    \caption{Class diagram representation of C1}
    \label{fig:c1}
\end{figure*}

We use case study C1 to illustrate how M2QCode models a hybrid quantum-classical system. Figure~\ref{fig:c1} presents the corresponding class diagram. The \textit{MachineLearningPDESolver} package represents the overall system architecture. The \textit{App} class serves as the application entry point, and the \textit{Presentation} package contains the \textit{MainForm} class, which defines the user interface for the input and configuration of the problem.

The \textit{BusinessLogic} package is responsible for formulating the optimization problem, while the \textit{ClassicalQuantumBusinessLogic} package handles the creation and coordination of the quantum optimization model. The \textit{QuantumLogic} package encapsulates the implementation of the Harrow–Hassidim–Lloyd (HHL) quantum algorithm and is associated with the quantum model. All quantum-related classes and packages are annotated with the UML stereotype $\langle\langle \textbf{Quantum} \rangle\rangle$, which distinguishes quantum components from classical ones.

\begin{figure}
    \centering
    \includegraphics[clip, trim=0 1530 0 0, width=0.37\textwidth]{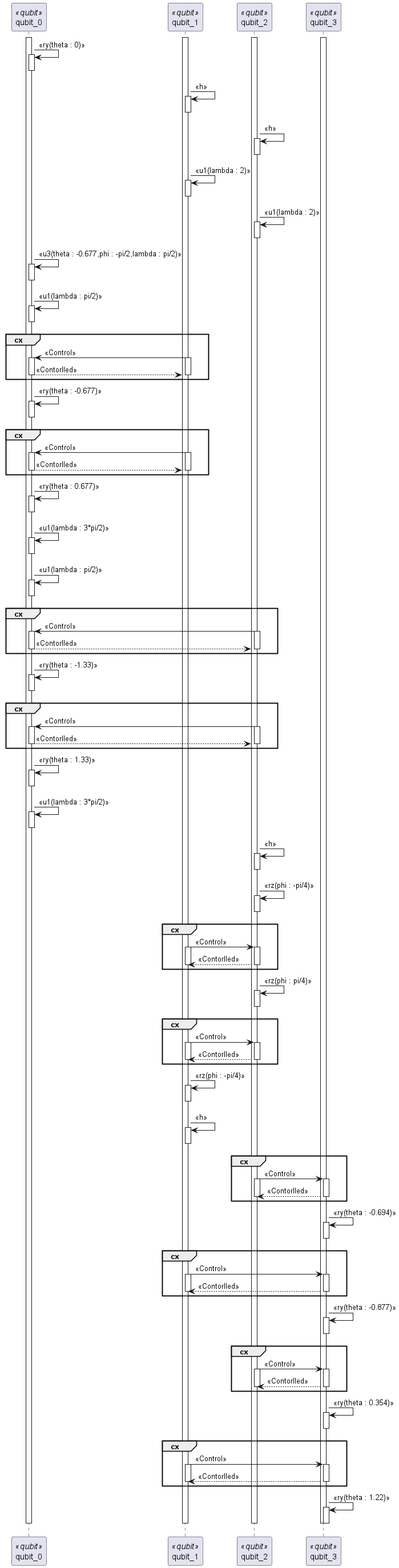}

    \vspace{0.1em}
    \begin{tikzpicture}
        \draw[decorate, decoration={zigzag, amplitude=1pt, segment length=4pt}] (-1,0) -- (7,0);
    \end{tikzpicture}
    \vspace{0.5em}

    \includegraphics[clip, trim=0 0 0 1470, width=0.37\textwidth]{figures/C1_copy.png}
    
    \caption{Sequence diagram of quantum algorithm in C1}
    \label{fig:c1_circuit}
\end{figure}

Figure~\ref{fig:c1_circuit} shows the sequence diagram of a 4-qubit HHL algorithm modeled using M2QCode. In this model, each \textit{Object} represents a qubit and is labeled with the $\langle\langle \textbf{qubit} \rangle\rangle$ stereotype. Gate operations are expressed as \textit{Message} elements: a self-message represents a single-qubit gate, with the message name indicating the specific quantum operation (e.g., \texttt{h}, \texttt{x}, \texttt{rz}).

For multi-qubit operations, a \textit{GroupMessage} is used, where the control qubit acts as the sender and the target qubit as the receiver. This abstraction effectively models controlled operations such as \texttt{CNOT}, preserving both structural and behavioral semantics of the circuit.

C1 includes both a class diagram and a sequence diagram, offering complementary perspectives. The class diagram provides a high-level architectural view of the system, while the sequence diagram captures the implementation details of the quantum algorithm, including the temporal ordering and interaction of qubits and operations.

\begin{tcolorbox}[size=title, rightrule=1mm, leftrule=1mm, toprule=0mm, bottomrule=0mm, arc=0pt, colback=gray!5, colframe=bleudefrance!75!black, breakable]
\textbf{Answer to RQ1:} M2QCode can model hybrid quantum-classical systems at both the structural (class) level and the quantum algorithm level, providing clear separation and integration of classical and quantum components.
\end{tcolorbox}

\subsubsection{RQ2: How effective is the generation of the quantum system?}

Table~\ref{tab:elements compare} presents the effectiveness of M2QCode in generating the classical part of the quantum-classical system from an information retrieval perspective. The results show an average precision of 82\% and a recall of 100\%. Interestingly, the number of generated elements exceeds the number of expected code segments.
Upon further analysis, we found that this is due to M2QCode’s inclusion of Python-specific language features, such as the automatic generation of default constructors for each class. These constructors are added to ensure correct object initialization. As shown in the table, the number of \#Operations in the generated code is equal to the sum of \#Operations and \#Classes in the UML model, supporting this interpretation. When these auto-generated constructors are excluded, the precision also reaches 100\%, aligning exactly with the expected output.
These findings demonstrate that M2QCode can effectively extract structural information from UML models and generate \textit{syntactically correct and semantically faithful} Python code.

\begin{table}[b]
\centering
\caption{Elements Completeness Comparison for Code Generation}
\label{tab:elements compare}
\resizebox{\linewidth}{!}{%
\begin{tabular}{@{}lcccccccc@{}}
\toprule
 &   & \textbf{C1} & \textbf{C2} & \textbf{C3} & \textbf{C4} & \textbf{C5} & \textbf{C6} & \textbf{C7} \\
\midrule
\multirow{4}{*}{\shortstack[l]{UML \\Class\\Diagram}} &
\#Packages   & 8  & 8  & 5  & 8  & 8  & 8  & 8   \\
 & \#Classes    & 13 & 10 & 9  & 9  & 9  & 12 & 10  \\
 & \#Operations & 10 & 21 & 11 & 11 & 11 & 20 & 11  \\
 & \#Attributes & 16 & 13 & 5  & 9  & 9  & 16 & 11  \\
\midrule
\multirow{4}{*}{\shortstack[l]{Python \\Code\\Generation}} &
\#Packages   & 8  & 8  & 5  & 8  & 8  & 8  & 8   \\
 & \#Classes    & 13 & 10 & 9  & 9  & 9  & 12 & 10  \\
 & \#Operations & 23 & 31 & 20 & 20 & 20 & 25 & 21  \\
 & \#Attributes & 16 & 13 & 5  & 9  & 9  & 16 & 11  \\
\midrule
\multirow{5}{*}{\shortstack[l]{RQ2: \\Classical \\Code\\ Effectiveness}} 
& \#Relevant elements   & 53 & 57 & 35 & 40 & 40 & 55 & 45  \\
& \#Irrelevant elements & 13 & 10 & 9  & 9  & 9  & 12 & 10 \\
& \#Missing elements    &  0 &  0 & 0  & 0  & 0  & 0  & 0   \\
& \#Precision           & 0.80 & 0.85 & 0.80 & 0.82 & 0.82 & 0.82 & 0.82  \\
& \#Recall              & 1.00 & 1.00 & 1.00 & 1.00 & 1.00 & 1.00 & 1.00  \\
\bottomrule
\end{tabular}%
}
\end{table}

Table~\ref{tab:quantum recall} shows that M2QCode achieved 100\% precision and recall in all QPLs and case studies, indicating that it can accurately identify quantum elements and generate complete quantum code segments.

\begin{table}[h!]
\centering
\caption{Precision and Recall of QPLs on Each Case}
\label{tab:quantum recall}
\resizebox{\linewidth}{!}{
\begin{tabular}{llccccccc}
\toprule
\textbf{Metric} & \textbf{QPL} & \textbf{C1} & \textbf{C2} & \textbf{C3} & \textbf{C4} & \textbf{C5} & \textbf{C6} & \textbf{C7} \\
\midrule
Precision & All QPLs & 1 & 1 & 1 & 1 & 1 & 1 & 1 \\
Recall    & All QPLs & 1 & 1 & 1 & 1 & 1 & 1 & 1 \\
\bottomrule
\end{tabular}
}
\end{table}

\begin{figure*}
    \centering
    \includegraphics[width=0.9\linewidth]{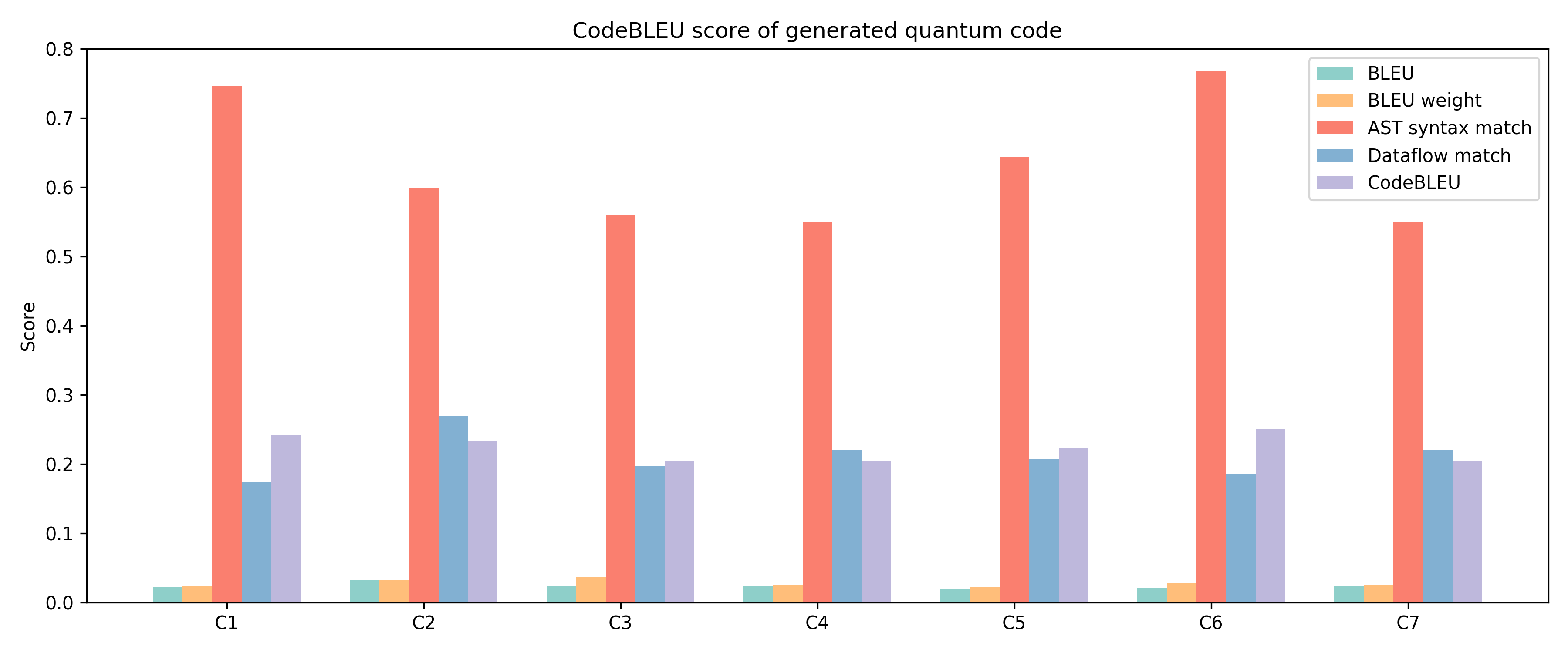}
    \caption{CodeBLEU score of generated Qiskit quantum code}
    \label{fig:codebleu}
\end{figure*}

Although M2QCode supports code generation for multiple QPLs, our analysis focuses on Qiskit, as the benchmark dataset provides reference implementations only in Qiskit. Figure~\ref{fig:codebleu} shows the CodeBLEU scores for each case.

M2QCode-generated programs achieve high syntax match scores based on AST comparisons. Even the lowest score (0.55 for C4) indicates a significant structural similarity to the reference circuits. However, the overall CodeBLEU scores are relatively modest.

A detailed analysis reveals that naming discrepancies have a significant impact on the BLEU and data flow components. M2QCode uses standardized identifiers (e.g., \texttt{qc}, \texttt{q}, \texttt{cRegister}), while the reference circuits use varied naming styles and often include execution setup such as circuit instantiation and measurement instructions, while also using ad hoc naming schemes and omitting boilerplate code. Although these textual mismatches are irrelevant to program semantics, they substantially impact BLEU and data flow scores.

\begin{figure}
    \centering
    \includegraphics[width=0.9\linewidth]{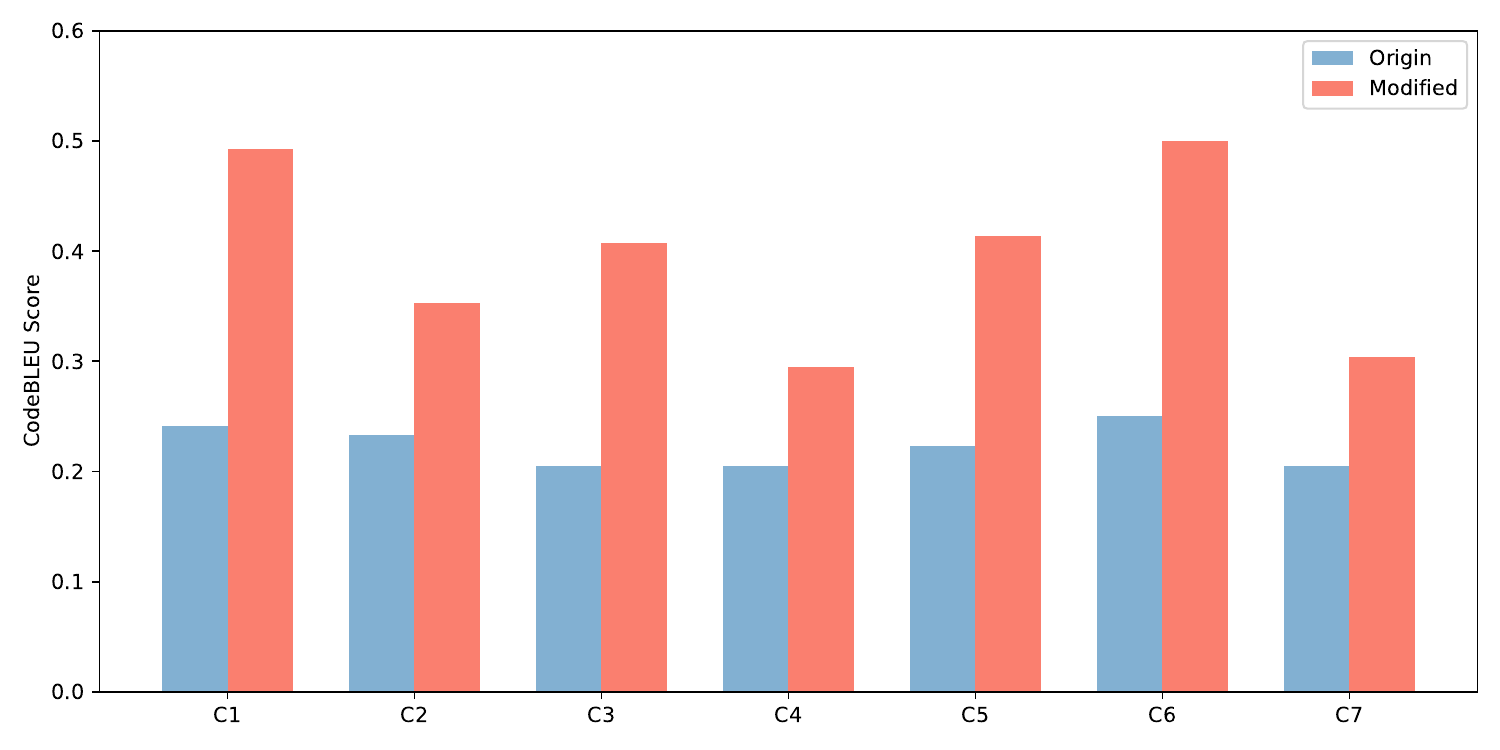}
    \caption{Comparison of CodeBLEU scores before and after variable renaming}
    \label{fig:codebleu compare}
\end{figure}

To validate this hypothesis, we manually adjusted the variable names in the generated code to match the reference implementations. As shown in Figure~\ref{fig:codebleu compare}, this alignment led to a significant improvement in CodeBLEU scores, confirming that M2QCode produces semantically and functionally equivalent code.

\begin{tcolorbox}[size=title,rightrule=1mm, leftrule=1mm, toprule=0mm, bottomrule=0mm, arc=0pt,colback=gray!5,colframe=bleudefrance!75!black,breakable]
\textbf{Answer to RQ2:} \black{M2QCode demonstrates that generated classical and quantum components preserve the structural and semantic information encoded in the original UML model.} It achieves high precision and recall, and the generated quantum code exhibits a high structural and semantic similarity to the reference implementations.
\end{tcolorbox}

\subsubsection{RQ3: Can M2QCode generate functionally equivalent quantum code across multiple QPLs?}

To evaluate the functional correctness of the generated quantum code, we measure the distance between the output distributions of the generated and reference circuits using the Kullback-Leibler (KL) divergence metric. Each program was executed 1024 times and the measurement results were used to approximate the output distribution.

To minimize the variation caused by different quantum execution environments, we used quantum simulators with comparable capabilities across all tested QPLs. This approach ensured that differences in output distributions were due solely to the M2QCode generation process, not back-end discrepancies.

Table~\ref{tab: kl compare} shows the KL divergence values between the generated and reference circuits. In all evaluated cases, the divergence remains below 0.1, indicating that the generated quantum programs closely approximate the expected output behavior. These results confirm that M2QCode can produce quantum programs functionally equivalent to reference implementations in different QPLs.

We note that for cases C1, C2, and C5, no results were obtained for the Q\# implementation. Further investigation revealed that these programs failed to execute due to using \texttt{u2} and \texttt{u3} gates, which are not natively supported in Q\#. These gates perform parameterized rotations around the x-, y-, and z-axes.

Although M2QCode provides callable placeholders for such composite gates, the developer must provide the internal implementations. After manually defining these gates using the available Q\# operations, all three programs executed successfully. This confirms that functional equivalence remains achievable in Q\# through manual adaptation when unsupported gates are involved.

\begin{tcolorbox}[size=title,rightrule=1mm, leftrule=1mm, toprule=0mm, bottomrule=0mm, arc=0pt,colback=gray!5,colframe=bleudefrance!75!black,breakable]
\textbf{Answer to RQ3:} M2QCode can generate functionally equivalent quantum code across multiple QPLs. The KL divergence results confirm that the generated programs closely match the expected output behavior, with low divergence values in all supported cases.
\end{tcolorbox}

\begin{table}[htbp]
  \centering
  \caption{Kullback-Leibler divergence between original and generated quantum circuits}
  \label{tab: kl compare}
  \resizebox{\linewidth}{!}{%
  \begin{tabular}{lcccccccc}
    \toprule
    \textbf{QPL} & \textbf{C1} & \textbf{C2} & \textbf{C3} & \textbf{C4} & \textbf{C5} & \textbf{C6} & \textbf{C7} & \textbf{Avg.} \\
    \midrule
    Qiskit  & 0.02 & 0.01 & 0.00 & 0.03 & 0.00 & 0.06 & 0.01 & 0.00 \\
    Cirq    & 0.01 & 0.01 & 0.00 & 0.03 & 0.00 & 0.07 & 0.02 & 0.00 \\
    Braket  & 0.03 & 0.01 & 0.00 & 0.02 & 0.00 & 0.06 & 0.02 & 0.00 \\
    Q\#     & --   & --   & 0.00 & 0.03 & --   & 0.04 & 0.02 & 0.00 \\
    \bottomrule
  \end{tabular}%
  }
\end{table}

\section{Discussion}
\label{sec:discussion}

\subsection{Interpretation of Experimental Results and Their Implications}

The evaluation results demonstrate the promising capabilities of M2QCode in modeling and generating hybrid quantum-classical systems. High precision and recall scores confirm the effectiveness of the framework in extracting structural and behavioral information from UML-based models. The results of CodeBLEU further support its ability to generate quantum code with strong syntactic and semantic similarity to reference implementations, while the low KL divergence values indicate functional correctness across different QPLs.

These findings highlight the feasibility of applying Model-Driven Development (MDD) to quantum software engineering. The ability to generate semantically equivalent quantum programs across multiple QPLs can significantly reduce development complexity, enhance code maintainability, and promote interoperability in multiplatform quantum software systems.

\subsection{Potential Improvements}

Although M2QCode shows strong capabilities, several areas remain for future enhancement.

\begin{itemize}
    \item \textit{Higher Abstract Level. M2QCode currently adopts a circuit-level modeling approach, allowing for the representation of a wide range of quantum circuits with fine-grained control. However, this representation remains relatively low-level, limiting abstraction. To address this, we are extending the framework to support modular modeling, allowing users to define and reuse custom circuit modules (e.g., Oracles in Grover’s algorithm) that frequently recur in quantum programs. This modular representation improves abstraction and reusability by reducing gate-level complexity.}

    \item \textit{Modeling Efficiency.} As quantum circuits grow in depth and complexity, manually modeling quantum algorithms using extended UML sequence diagrams becomes increasingly labor-intensive. Future work should explore abstractions or automation techniques to reduce the human effort required to model large-scale quantum algorithms.

    \item \textit{User-Defined Gate Composition.} As observed in the Q\# experiments, gates such as \texttt{u2} and \texttt{u3} are not natively supported and require manual implementation. Supporting user-defined composite gates would allow developers to define reusable gate blocks, improving both modeling efficiency and compatibility across QPLs.

    \item \textit{Scalability to Quantum Machine Learning.} Quantum machine learning (QML) is widely applied in domains such as chemistry, physics, and materials science. For users without deep experience in quantum programming, developing QML applications across platforms can be challenging. Extending M2QCode to support QML frameworks, such as PennyLane and TensorFlow Quantum, would make it more accessible to a broader audience.
\end{itemize}

\section{Related Work}
\label{sec:related work}

\subsection{Classical Code Generators}
Model-Driven Development (MDD) has been widely used in classical software engineering to bridge the gap between high-level models and executable code. Classical code generators~\cite{kelly2008domain, usman2008ujector, syriani2018systematic} typically rely on modeling languages, such as UML and domain-specific languages (DSLs), to support automated transformations into general-purpose programming languages, such as Java, C++, and Python. Tools such as Eclipse Acceleo and Papyrus enable model-to-text (M2T) transformations~\cite{czarnecki2006feature} and employ Platform Independent Models (PIMs) and Platform Specific Models (PSMs) to manage different abstraction levels.

Although effective for classical systems, these tools are not designed to handle the unique features of quantum software, such as qubits, quantum gate operations, and hybrid classical-quantum interactions. This limitation highlights the need for MDD techniques specifically tailored to the quantum domain, which motivates the development of quantum code generation frameworks.

\subsection{Quantum Code Generators}
Several quantum code generation approaches have been proposed in recent years, including both visual modeling tools and AI-assisted solutions. For example, Jimenez-Navajas {\it et al.} proposed a quantum system generator based on a quantum UML Profile~\cite{jimenez2025code}, which extends classical UML with quantum-specific stereotypes to support the modeling of qubits, gates, and measurements. Although effective in generating hybrid systems, this approach is tightly coupled with Qiskit and lacks support for multiplatform code generation and some widely used quantum gates (e.g., \texttt{CSWAP}).

Another line of work is represented by the Qiskit Code Assistant~\cite{dupuis2024qiskit}, which leverages large language models (LLMs) fine-tuned for quantum programming to generate code from natural language prompts. Although promising, these tools do not provide guarantees of program correctness and lack the structured model-based generation capabilities offered by MDD.

Classiq~\cite{minerbi2022quantum} adopts a proprietary high-level modeling language, Qmod, which serves as an abstract intermediate representation and is compiled into Qiskit programs through a custom backend compiler. 
However, Qmod is not aligned with established methodologies in the Model-Driven Development (MDD) community or standard practices in quantum system modeling, and the Classiq framework also lacks support for quantum programming languages beyond Qiskit. 
These limitations restrict its applicability in multiplatform quantum software engineering, particularly in industrial contexts where automated cross-QPL code generation and deployment are essential.

Some existing quantum code generation frameworks~\cite{gemeinhardt2024model, gemeinhardt2023model} based on Model-Driven Development (MDD) abstract quantum circuits at the layer level, grouping operations into predefined structural blocks. 
While this approach improves abstraction, it imposes modeling constraints: users must define layer structures in advance, which reduces flexibility and limits expressiveness. 
Consequently, such frameworks are suitable only for a restricted class of quantum programs and generally do not support multiple quantum programming languages or hybrid quantum-classical modeling, further limiting their applicability in industrial quantum software engineering.

These limitations leave a gap for frameworks such as M2QCode, which aim to integrate model-driven design with platform-agnostic quantum program generation, supporting multiple QPLs, including Qiskit, Cirq, Q\#, and Braket.

\section{Conclusion}
\label{sec:conclusion}

This paper presented \textbf{M2QCode}, a Model-Driven Development (MDD) framework for automatically generating multiplatform quantum programs. 
M2QCode addresses key challenges in the NTT NVQS development pipeline and supports scalable quantum software development across multiple quantum programming languages (QPLs), including Qiskit, Cirq, Q\#, and Braket.
The framework enables system-level modeling of hybrid quantum-classical systems using an extended UML-based specification and supports automated transformation into executable code for multiple QPLs.

We evaluated M2QCode through case studies involving seven real-world quantum systems, demonstrating its ability to model hybrid systems, extract structured representations, and generate syntactically correct and functionally equivalent code.

By integrating system modeling with code generation, M2QCode supports model-driven development for quantum software, reducing design complexity and enabling broader platform interoperability. Although the framework has shown promising results, there are opportunities for further enhancement. One such direction is the addition of user-defined gate composition support. This would address the limitations identified in our case studies, such as the lack of native support for certain gates in Q\#, and would facilitate the reuse of common structural patterns in quantum algorithms. Allowing users to define composite gate functions would reduce modeling effort and improve scalability for larger systems.

\newpage
\bibliographystyle{IEEEtran}
\bibliography{qse-bibliography}





\end{document}

%% file: macros.tex
\usepackage{times}
\usepackage{graphicx}
\usepackage{epsf}
\usepackage{verbatim}
\usepackage{url}
\usepackage{color}
\usepackage{alltt}

\newcommand{\SmallSpace}{\vspace*{-1.4ex}}

